\documentclass[aps,prb,showpacs,showkeys,nofootinbib,reprint,
superscriptaddress,preprintnumbers]{revtex4-1}
\usepackage{amsmath,amssymb,bm,mathrsfs}
\usepackage{srcltx}
\usepackage{dsfont}
\usepackage{epsfig}
\usepackage{slashed}
\usepackage{bbold}
\usepackage{psfrag}
\usepackage{xcolor}
\PassOptionsToPackage{caption=false}{subfig}
\usepackage{subfig}
\usepackage{xfrac}
\usepackage{multirow}
\usepackage{booktabs}
\usepackage{hyperref}
\hypersetup{colorlinks=true,linkcolor=violet,citecolor=violet,urlcolor=violet}
\usepackage{relsize}
\usepackage{verbatim}
\usepackage{natbib}
\usepackage{float}
\usepackage{graphicx}

\def\({\left(}
\def\){\right)}

\usepackage{soul}

\newcommand{\B}{\text{B} }
\newcommand{\T}{\text{T} }

\begin{document}
\title{ Multi-band character revealed from  Weak-antilocalization effect in Platinum thin films} 
\author{Subhadip Jana$~$}
\email{subhadip.j@iopb.res.in}
\affiliation{Institute of Physics, Bhubaneswar, 751005, India}
\affiliation{Homi Bhabha National Institute, AnushaktiNagar, Mumbai, 400094, India}
\author{$~$T. Senapati}
\affiliation{School of Physical Sciences, National Institute of Science Education and Research, Bhubaneswar, 752050, India}
\affiliation{Homi Bhabha National Institute, AnushaktiNagar, Mumbai, 400094, India}
\author{K. Senapati}
\affiliation{School of Physical Sciences, National Institute of Science Education and Research, Bhubaneswar, 752050, India}
\affiliation{Homi Bhabha National Institute, AnushaktiNagar, Mumbai, 400094, India}
\author{D. Samal}
\email{dsamal@iopb.res.in}
\affiliation{Institute of Physics, Bhubaneswar, 751005, India}
\affiliation{Homi Bhabha National Institute, AnushaktiNagar, Mumbai, 400094, India}
\begin{abstract}
Platinum (Pt) has been very much used for spin-charge conversion  in spintronics research due to it's large intrinsic spin-orbit interaction. Magnetoconductance originated from weak-antilocalization  effect in quantum interference regime is used as a powerful tool to obtain  the microscopic information  of spin-orbit interaction  and  coherence phase breaking scattering  process among itinerant electrons. To acquire the knowledge of different types of scattering processes, we have performed  magnetoconductance study  on  Pt thin films which manifests multi-band (multi-channel) conduction.  An extensive analysis of quantum interference originated weak-antilocalization effect  reveals the existence of strong (weak) inter-band scattering between two similar (different) orbitals.   Coherence phase breaking lengths ($l_{\phi}$)  and their temperature dependence  are found to be significantly different for these two conducting bands. The observed effects are consistent with theoretical predication that there exist three Fermi-sheets with one $s$ and two $d$ orbital character. This study provides the evidence of  two independent non-similar conducting channels and presence of  anisotropic spin-orbit interaction along with  $e$-$e$ correlation  in Pt thin films.    
\end{abstract}
\maketitle
\section{Introduction}
Gaining control over  electron spin degree of freedom is very much desirable in the field of spintronic research \cite{RevModPhys.80.1517,RevModPhys.76.323}. 
In recent days, spin-orbit interaction (SOI)  has been found to provide promising strategy for electrical manipulation spin/magnetism in spintronic devices \cite{datta1990electronic,adma,spintronics,spintronic_1}. This includes the creation of spin current from transverse charge current effect (SHE) \cite{d1971possibility,dyakonov1971current,RevModPhys.87.1213}, and the exertion of a torque on a local magnetization from electrical current by spin-orbit torque effect \cite{zhu2021maximizing,han2021materials,song2021spin,RevModPhys.91.035004}. At a microscopic level, the asymmetric  spin dependent electron scattering induced by SOI lies at the heart of the above phenomena \cite{valenzuela2006direct,PhysRevLett.94.047204}. Besides, a broader implication of SOI is realized in the design of topological materials with their potential use in low energy dissipation and faster magnetization switching \cite{spintronics_topo,puebla,spintronic_switch,liu_spin}. Materials with high Z element (Z is atomic number)  are ideal candidates to look for spin-orbit interaction induced phenomena (SOI strength $\propto $Z$^{4}~~$ \cite{PhysRevLett.41.805}).  In particular, 5$d$ transition metal  Pt  has drawn lot of attention for spin-charge conversions and spin-torque effect in Pt/magnetic layer based hetrostructures  due to its intrinsic high SOI \cite{hetro_Pt_1,hetro_Pt_2,hetro_Pt_3,hetro_Pt_4}. 
\textcolor{black}{Further, the  observation of  inverse spin hall effect (ISHE) (reverse process of SHE, $i.e.$ creation of charge current from transverse spin current) in Ni$_{81}$Fe$_{19}$/Pt and in Pt wire at room temperature provided an effective way to  detect spin current \cite{Pt_ISHE_sato,PhysRevLett.98.156601}.} 
In view of  the extensive use of Pt in spintronics owing to its chemical inertness, easy fabrication of device and most importantly strong intrinsic spin-orbit interaction, it is important to have a comprehensive understanding of the effect of spin-orbit interaction on electronic transport in bare  metallic Pt thin films. 


%
%
%

Quantum interference originated weak-localization (WL) or weak-antilocalization (WAL) effect is very much sensitive to the  spin-orbit interaction (SOI) scattering process in conducting systems \cite{bergmann_rev}.  In real system,  any type of deviation from perfect crystalline  structure acts as a source of scattering potential for itinerant electrons.  Scattered electrons propagating along time-reversed, identical self-intersecting trajectories  known as ``Cooperon loop" (CL) \cite{romer,colman_book_1} interfere constructively/ destructively to give rise to suppression (WL)/ enhancement (WAL)  of conductivity. 
%
Experimentally, WL/WAL is usually determined from conductance correction  to classical Drude conductivity   at low  temperature and in presence of external magnetic  field. The interference correction tends to vanish for most of  trajectories after averaging over the random scattering potentials, except for those scattered electrons which propagate in CL.

%
The application of uniform external magnetic field breaks the time  reversal symmetry required for the interference effect  and   induces an additional relative phase shift (due to enclose magnetic flux) between the two electrons traversing CL . This effect  suppress constructive/destructive interference, as a result  conductance gets enhanced (positive)/decrease (negative)  with application of  magnetic field in the WL/WAL regime \cite{BERGMANN19841,PhysRevLett.103.226801}. Therefore, variation of  magnetoconductance is used as a sensitive probe to detect quantum interference effect.
%
Traditionally, the manifestation of  WAL has been attributed to SOI in material. In order for WAL to occur, it is crucial to have a $\pi$ phase shift between two electron trajectories. 
The rotation of  spin $1/2$ particle    by $4\pi$ is equivalent to the identity operation. 
However, for Pt thin films  $2\pi$ rotation of itinerant electron spin induced by SOI gives rise to  $\pi$ phase shift  resulting in WAL effect \cite{PhysRevLett.48.1046,mccann2009staying}.  

In this work, we have  carried out an indepth magnetotransport study to examine the underlying SOI scattering of itinerant electrons from    WAL effect in Pt thin films. It has been realized that SOI scattering strength (proportional to the characteristic spin-orbit magnetic field ($B_{so}\sim 2$ T))   is very much stronger than coherent phase breaking scattering strength   ($B_{\phi}\sim 0.004$ T at 2 K). The corresponding $B_{so}$ and $B_{\phi}$ are extracted by using Hikami-Larkin-Nagaoka equation (Eq.\ref{Eq_HLN_multi}).  Further, it has been found that the single conduction channel can not be fitted well with experimental data, rather two independent conduction channels need to be considered  to reconcile with experimental data.   
\textcolor{black}{To get more information about channels and their orbital symmetry, we examined temperature dependent behavior of $B_{\phi}$.
It is found that $B^{1}_{\phi}(T)$ for one channel exhibits  prominent temperature dependence whereas $B^{2}_{\phi}(T)$ corresponding to other channel shows weak temperature dependence. The observed  difference in the temperature variation of $B^{1}_{\phi}(T)$ and $B^{2}_{\phi}(T)$  are attributed to two conducting channels made of  orbitals with different symmetry. The band  originated from more symmetric orbitals  are less sensitive to disorder whereas the band  originated from anisotropic orbitals are more sensitive to disorder and it is reflected in temperature dependent $B^{1}_{\phi}(T),B^{2}_{\phi}(T)$.   The observed effects are consistent with theoretical predication that there exist three Fermi-sheets (FS) with one $s$ and two $d$ orbital character \cite{Pt_dft,PhysRevB.2.4813,PhysRevLett.100.096401}. This study provides the evidence of  two independent nonsimilar conducting channels (one channel is made of $s$ orbital FS and other is originated from combining two $d$ orbital FS (illustrated in Fig.\ref{fig_schematic})) and presence of  anisotropic spin-orbit interaction along with  $e$-$e$ correlation  in Pt thin films.}


\section{$\text{Experiment}$}

Platinum thin films were grown on Si/SiO$_{2}$ substrates using  DC magnetron sputtering at room temperature with a base pressure $5\times 10^{-8}$ mBar. Before deposition, photolithography patterning with a Hall bar geometry was done with a mask aligner MDA-400M-N on $5\times5$  mm$^{2}$ substrates. The  thickness variation of Pt layer was achieved from a  single deposition-run by rotating substrate plate holder. The structural characterization was carried out using a high-resolution X-ray diffractometer. The films were found to be  polycrystalline with grains  mainly  grown/oriented  on $(111)$ plane. The thickness of the films was determined  to be  16 nm and 23 nm with a roughness  about  3 \AA $~$ from X-ray reflectivity (XRR) measurement. The magneto transport measurements were performed in a Cryogenic physical property measurement system (PPMS) with magnetic fields applied parallel and perpendicular to the film surface. We used Keithley 6221 as the current source and a Keithley 2182A nano voltmeter for better data resolution.


\section{$ \text{Resistivity at low temperature}$} 
\label{rho_T}
\begin{figure}[t!]
	\begin{center}
		\includegraphics[width=9 cm,height=6 cm]{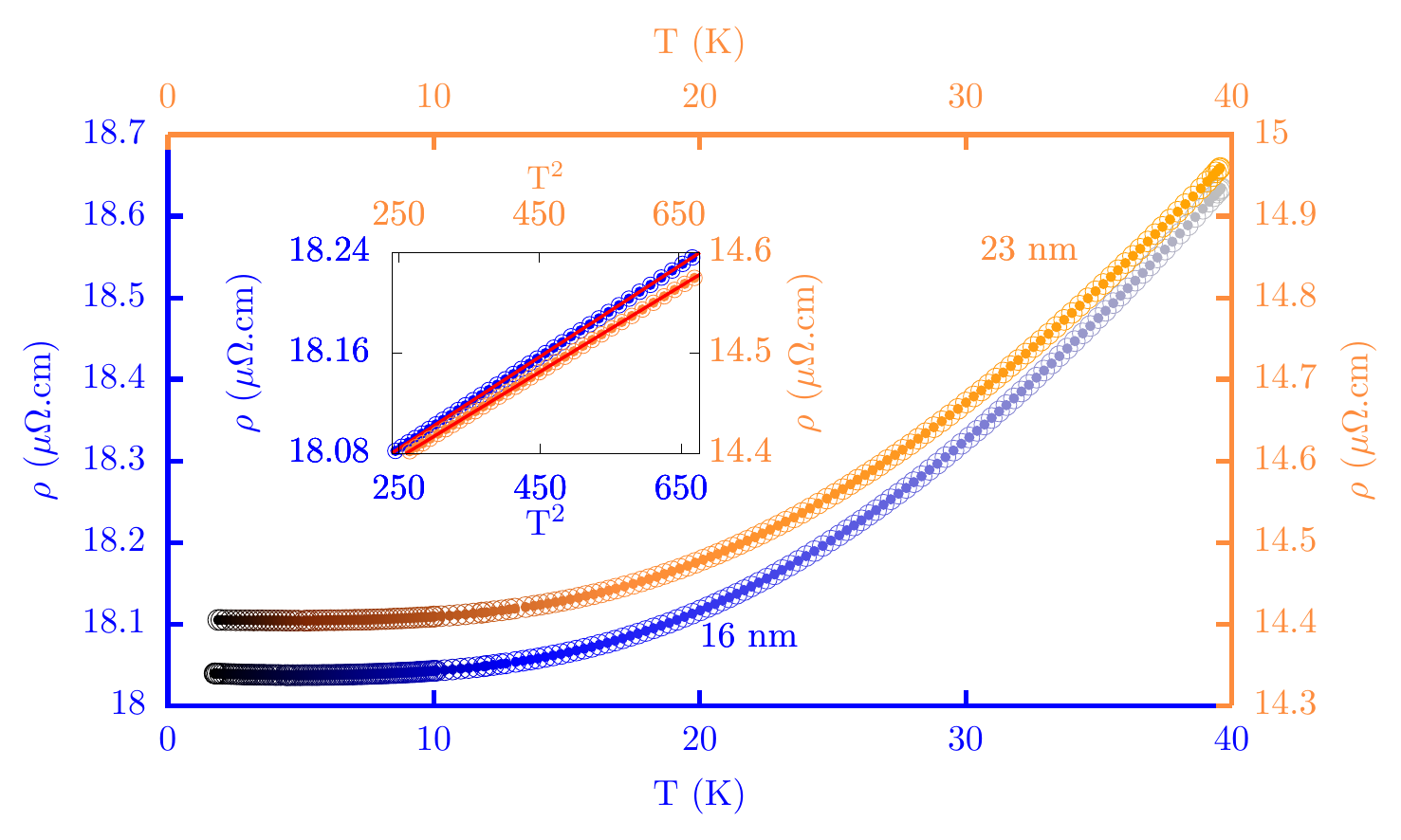}\\
	\end{center}
	\caption{\small \textcolor{black}{{Resistivity$~\rho(T)~vs~T$  plot for  Pt 16 nm (blue color) and  23 nm (orange color) films, darken color at low temperature indicates quantum interference  regime.  EEI originated   $\rho(T) ~\propto ~ T^{2}$ in the temperature range 15 K $<~T~<$ 26 K is illustrated in inset  (solid red color lines represent fitting of  Eq.(\ref{fermi_liq_eq})).}}
		\label{fig_rho_T}}
\end{figure}



The temperature dependent resistivity ($\rho$) for 16, 23 nm thick films (  Fig.\ref{fig_rho_T}) exhibits positive temperature coefficient ($\frac{d\rho}{dT} >0$) above 6 K  indicating metallic character. 
Moreover, in the intermediate temperature range 15 K $<~T~<$ 26 K, the 
resistivity follows  quadratic temperature  dependence which is attributed to $e$-$e$ coulomb interaction (EEI) 
  \cite{Pt_T2,PhysRevLett.20.1439,jacko2009unified} and is shown in inset of Fig.\ref{fig_rho_T}. Experimental data in the range temperature 15 K $<~T~<$ 26 K the  fits well with Eq.(\ref{fermi_liq_eq}), 
\begin{equation}
\rho(T) = \rho_{0} + AT^{2},
\label{fermi_liq_eq}
\end{equation}
where $\rho_{0}$ is residual resistivity and $A$ accounts for EEI contribution. 
From fitting, we obtain  $A$ equal to $3.6\times 10^{-4} ~~\mu . \Omega.$cm.K$^{-2}$ and $~3.36 \times 10^{-4} ~~\mu . \Omega.$cm.K$^{-2}$ for 16 nm and 23 nm thick films respectively. 
The obtained value of $A$ is higher for 16 nm film  as compared to 23 nm and it is consistent with  theory that $A~\propto ~ E^{-2}_{F} ~ \propto ~ (m^{*})^{2}$, where $E_{F}$ and $m^{*}$ are Fermi energy, effective mass of electron respectively \cite{T2_science}.    
The  extracted values of $A$ are slightly higher than the reported value of ($A \sim 10^{-5}~\mu.\Omega.$cm.K$^{-2}$) for bulk Pt \cite{jacko2009unified} which could be due to  reduction of system dimension.  

In low temperature regime (2 K$<T<6$ K), $\rho(T)$ exhibits an upturn which is attributed to the combined effect of quantum interference (WL/WAL) and EEI correction  in quasi 2D limit. This regime is denoted by dark color shed in Fig.\ref{fig_rho_T} and it is discussed in Sec:\ref{WAL_T} in terms of sheet conductance correction.          
\section{$\text{Signature of WAL Effect  from Temperature}$\\ $\text{ dependent sheet conductance} $}
\label{WAL_T}
\begin{figure}[t!]
	\begin{center}
		\includegraphics[width=8.4 cm,height=7.6 cm]{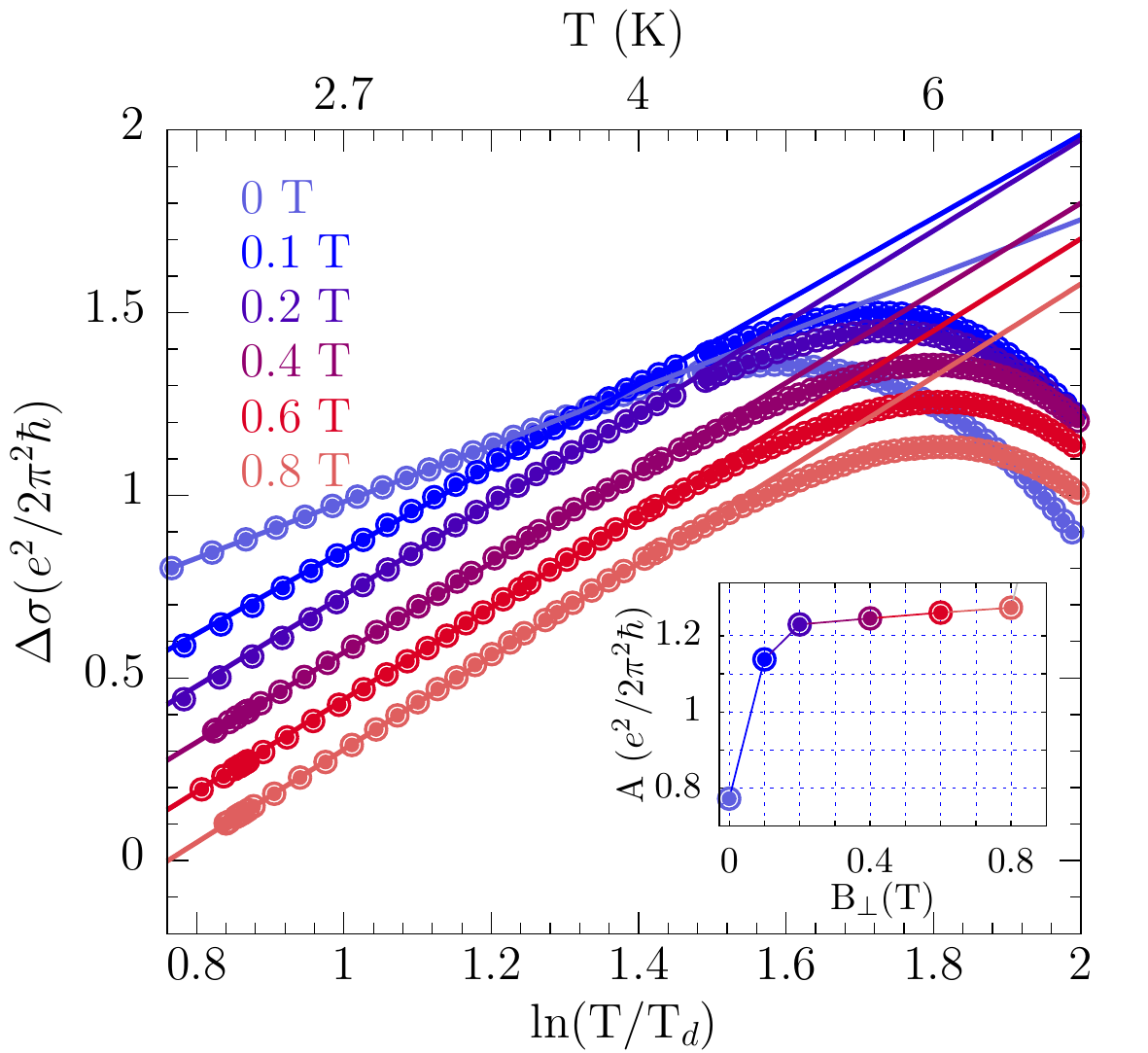}
	\end{center}
	\caption{\small{Sheet conductance (relative change) $~\Delta \sigma(T)~vs~\ln(T/T_{d})$ is plotted at different constant perpendicular magnetic fields ($0 \leq \B_{\perp} \leq 0.8 ~$T)  for Pt 16 nm thick film. Solid lines represent the fitting of $~\Delta \sigma(T)~vs~\ln(T/T_{d})$ using Eq.(\ref{eq_sigma_Total}) in the low temperature regime. (The upper x-axis denotes the real temperature and T$_{d} = 1$ K $~$ \cite{Td}) (vertically	offset to plots are given for clarity). The extracted coefficient $A$ for different $\B_{\perp}$ is shown in the inset.}
		\label{fig_sigma_ln_T}}
\end{figure}


At low temperature due to weak thermal agitation, electrons  are able to move longer distance (on an average)  in it's trajectory with maintaining   phase coherence which is known as phase coherence length ($l_{\phi}$).  Quantum interference manifests prominently when phase coherence length is  very much longer than mean free path ($l_{e}$) of electrons ($l_{\phi} >> l_{e}$). One of the indications of quantum interference correction is  $\ln(T)$ dependence of sheet conductance ($\Delta \sigma_{QI}$) in 2D limit ($ l_{\phi} > t$, where $t$ is film thickness) and in general, for $N$ independent  conducting channels or Fermi sheets,  $\Delta \sigma_{QI}$ can be expressed as \cite{hikami1980spin,tv_rev,PhysRevB.86.205302},
\begin{equation}
\label{eq_sigma_T_QI}
\Delta \sigma_{QI}(T) =  N\gamma_{int}\alpha p \frac{e^{2}}{2 \pi^{2} \hbar}\ln(T/T_{0})
\end{equation}
where $\gamma_{int}$ is related with interaction coupling strength among independent Fermi sheets (strong coupling and zero coupling lead to $N\gamma_{int} \sim 1$ and $N\gamma_{int} \sim N$ respectively),  $p$ is related with temperature exponent of phase coherence length ($ l^{2}_{\phi} ~ \propto ~ T^{-p} ~$\cite{Altshuler_1982,PhysRevB.27.5976,PhysRevB.65.180202}), $T_{0}$ depends on $l_{e}$  and  $\alpha$ takes different values depending upon  dominant scattering process involved. For three extreme situations $\alpha$  follows as: $\alpha \sim -1/2$ in strong spin-orbit scattering with absence of magnetic impurity scattering (WAL), $\alpha \sim 1$ in quantum coherent regime ($l_{\phi} > l_{e}$) in absence of spin-orbit and  magnetic impurity scattering (WL), $\alpha \sim 0$ in strong magnetic impurity scattering (which drives towards  classical scenario)\cite{hikami1980spin}. Therefore, the coefficient of $\ln(T)$ in Eq.(\ref{eq_sigma_T_QI}), $A_{QI} =  N\gamma_{int}\alpha p \frac{e^{2}}{2 \pi^{2} \hbar}$ contains  crucial information about microscopic scattering process.
Further, EEI contribution to sheet conductance ($\Delta \sigma_{e}$) exhibits also similar $\ln(T)$ dependence. For $N$ independent channels  in 2D limit,  it can be expressed as \cite{PhysRevLett.44.1288,fuku_1,fuku_2,fuku_3,dirac_2},
\begin{equation}
\label{eq_sigma_T_EEI}
\Delta \sigma_{e}(T) = N \frac{e^{2}}{2 \pi^{2} \hbar} (2-2F)\ln(T/T_{e})
\end{equation}
where $F$ is averaged screened Coulomb interaction  over Fermi surface, normalized with zero momentum transfer in EEI scattering process.   
Therefore, total correction to ($\sigma_{total}$) can be expressed as, 
\begin{equation}
\label{eq_sigma_Total}
\Delta \sigma_{total}(T) = A \frac{e^{2}}{2 \pi^{2} \hbar} \ln(T/T^{'})
\end{equation}
where $A = N( \gamma_{int} \alpha p + (2-2F))$ and $T^{'}$ is a constant.

Hence, it is not  straight froward to extract the value of $\alpha$ from zero magnetic field $\Delta \sigma(T)~vs \ln(T)$ experimental data, when the contribution from quantum interference and EEI coexist.  To overcome this issue,
one needs to  exploit  the temperature dependence of sheet conductance at constant weak magnetic fields since it will suppress mainly quantum interference  contribution.  

Quantum interference effect (WL/WAL)  originates from  particle-particle channel interaction  between two electrons traversing in ``Cooperon loop" (CL)  and it is very much sensitive to enclosed magnetic flux through the CL due to applied external magnetic field. However,  EEI effect  is governed by  particle-hole channel interaction  and it can not be influenced by external weak  magnetic field \cite{PhysRevLett.44.1288}(EEI can be influenced in higher magnetic field if $g\mu_{B}~B> 1/\tau_{so}$, where $\tau_{so},~g$  and $~\mu_{B} $ are SOI scattering time, Lande-g factor and Bhor-magneton respectively) \cite{tv_rev,PhysRevB.22.5142,PhysRevB.26.4009,e_e_logB_1,e_e_logB,dirac_2}. 
As a consequence of the above  fundamental differences between  quantum interference effect (WL/WAL) and EEI,  one can disentangle WL/WAL and EEI contribution  by applying (perpendicular to film surface) constant  weak external   magnetic fields. The applied field will suppress WL/WAL contribution, while  EEI effect remains unaffected.  Hence, the  change in  $\ln(T) $ coefficient upon the application of weak magnetic field will provide only WL/WAL contribution ($i.e. ~ A_{QI}$ in Eq.(\ref{eq_sigma_T_QI})).     

We investigated the variation of $\ln(T)$ coefficient ($A$) in $\sigma(T)$ with  applying constant perpendicular external magnetic fields ranging from 0 - 0.8 T,  shown in Fig.\ref{fig_sigma_ln_T}.  A systematic change in  $A$ was observed with the increment of external constant magnetic field and we obtained maximum change in $\ln(T)$ coefficient  as $ A_{QI} = A|_{B=0} - A|_{B=0.8}  = -0.56$ (in quantum conductance unit, $e^{2}/2\pi^{2}\hbar$). 
To evaluate the value of $\alpha$ from $A_{QI}$ that contains the information about dominant scattering process, one requires to know the values of  $ N\gamma_{int}$ and $~p$.  
The $p$ adapts a universal value depending upon the  dominant interaction responsible for inelastic scattering process in the system. In particular, for EEI originated inelastic scattering, $l^{-2}_{\phi}~(l^{-2}_{\phi}\propto T^{p})$ exhibits linear temperature dependence at low-temperature and it shows a crossover from $T$ to $T^{2}\ln(T)$ behavior with the increment of temperature \cite{PhysRevB.65.180202}.

Two possible cases can be invoked  to assess the value of $\alpha$:  
(i) For Pt thin films,  resistivity follows quadratic temperature dependence (14 K $<T<$ 23 K) (Sec:\ref{rho_T}) which  signifies the  dominance of electron-electron scattering over electron-phonon scattering and gives rise to $p=1$ at lower temperature. Considering single channel $N=1$ or strong coupling interaction $N\gamma_{int} \sim 1$,  one can obtain $\alpha \sim  -1/2$ (Eq.\ref{eq_sigma_T_QI}).
(ii) Effective two independent channels $N=2$ and $p=1$  (extracted from magnetoconductance analysis, Sec:\ref{B_perp})  give rise to  $\alpha \sim -0.28$ (Eq.\ref{eq_sigma_T_QI}). 


\textcolor{black}{From detailed magnetotransport analysis (discussed in Sec:\ref{WAL}), it is realized that the conduction takes place through two independent channels $N=2$  in Pt thin films. Considering two channel conduction, Eq.\ref{eq_sigma_T_QI} provides the value of $\alpha=-0.56/(N\gamma_{int}p) =-0.56/(2\times1) \sim -0.28$ (as discussed above) which is very much unexpected for high spin-orbit coupled system. To find out the source of this discrepancy, we re-looked into how  the explicit temperature dependence arises in the sheet conductance correction (Eq.\ref{eq_sigma_T_QI}) due to quantum interference effect. One can primarily visualize that with the increment of temperature, random thermal agitation enhances the inelastic scattering among itinerant electrons and as a consequence, $l_{\phi}$ becomes temperature dependent (explicit functional form is determined by the dominating scattering mechanism ($e$-$e$, $e$-phonon scattering) at non zero temperature). The $l_{\phi}$ fixes the upper cut-off spatial dimension for quantum interference effect and the interference originated correction is proportional to the available coherence area ($\propto l^{2}_{\phi}$).
Thus, for single channel with phase coherence length $l_{\phi}$, $\Delta \sigma_{QI} \propto -\ln (l^{2}_{\phi}/ l^{2}_{e})$.        
On the contrary, dominant impurity induced inelastic scattering leads to temperature independent $l_{\phi}$ \cite{PhysRevB.84.161405} and such kind of conducting channel can not contribute to $\Delta \sigma_{QI} (T)$ considerably.}
%
 Therefore, for a system with $N$ non-similar independent conduction channel, $\Delta \sigma_{QI} (T)$ can  be expressed as,
\begin{equation}
\label{eq_sigma_T_phase_N}
\Delta \sigma_{QI}(T) =   -\alpha  \sum_{n=1}^{N} \frac{e^{2}}{2 \pi^{2} \hbar}\ln((l^{n}_{\phi}/l^{n}_{e})^{2}). 
\end{equation}
%
%
It has been observed from magnetoconductance analysis that the examined Pt thin films possesses two independent conducting channels, one of which  shows prominent temperature dependence ($B^{1}_{\phi}(T) \propto T$, Fig.\ref{fig_16nm_sigma_HLN_frac}) at lower temperature regime (2 K $<T<6$ K), and   $B^{2}_{\phi}(T)$ associated with other channel follows a negligible variation with  temperature. This implies that contribution to the  slope of $\ln(T)$ (Eq.\ref{eq_sigma_T_phase_N}) from second channel is negligible. Thus, Eq.\ref{eq_sigma_T_phase_N} can effectively be converted into
\begin{equation}%
\begin{split}
\Delta \sigma_{QI}(T) &=  -\alpha  \frac{e^{2}}{2 \pi^{2} \hbar}\Big( \ln(l^{1}_{\phi}/l^{1}_{e})^{2} + \ln(l^{2}_{\phi}/l^{2}_{e})^{2} \Big) \\
& =  -\alpha  \frac{e^{2}}{2 \pi^{2} \hbar}\Big( \ln(l^{1}_{\phi}(T)/l^{1}_{e})^{2} + \ln(l^{2}_{\phi}(T)/l^{2}_{e})^{2} \Big) \\ 
& \sim  \alpha  \frac{e^{2}}{2 \pi^{2} \hbar}\Big( \ln(T/T^{1}_{0}) + c \Big)
\end{split}
\label{Eq_ln_T_multi}
\end{equation}
where, $c$ is a temperature independent constant as $l^{2}_{\phi}$ exhibits negligible temperature dependence. Since only one channel contributes  logarithmic temperature correction to $\Delta \sigma_{QI}(T)$, we obtained $\alpha=-0.56/(N\gamma_{int}p) =-0.56/(1\times1)\sim -1/2 $ which is consistent with theoretical value of $\alpha$ in presence of strong spin-orbit scattering.

We now turn into the estimation EEI strength which is related with screened Coulomb potential $F$. For strong interaction, $F$ is small fractional number  and for free electrons $F \sim 1$.  $F$ can be evaluated by considering that at a magnetic field of 0.8 T, the quantum interference effect is completely exhausted and then one can approximate the coefficient $A|_{B=0.8 \T} \sim  N\times 2(1-F)=1.22$, ($N=2$ as discussed before) which leads to $F \sim 0.7 $. The  5$d$ correlated transition metal Pt exhibits  higher value of $F$ which   indicates   $e$-$e$ Coulomb interaction is  weaker in comparison with  strongly electron correlated 3$d$ transition metal Cu $F_{Cu} \sim 0.5$ \cite{PhysRevB.33.6631}.   
\section{$\text{ WAL Effect  from Magnetic field Variation}$}
\label{WAL}
\subsection{Perpendicular Magnetic field ($\B_{\perp}$)}
\label{B_perp}
\begin{figure}[h!]
	\begin{center}
	\includegraphics[width=7cm,height=7 cm]{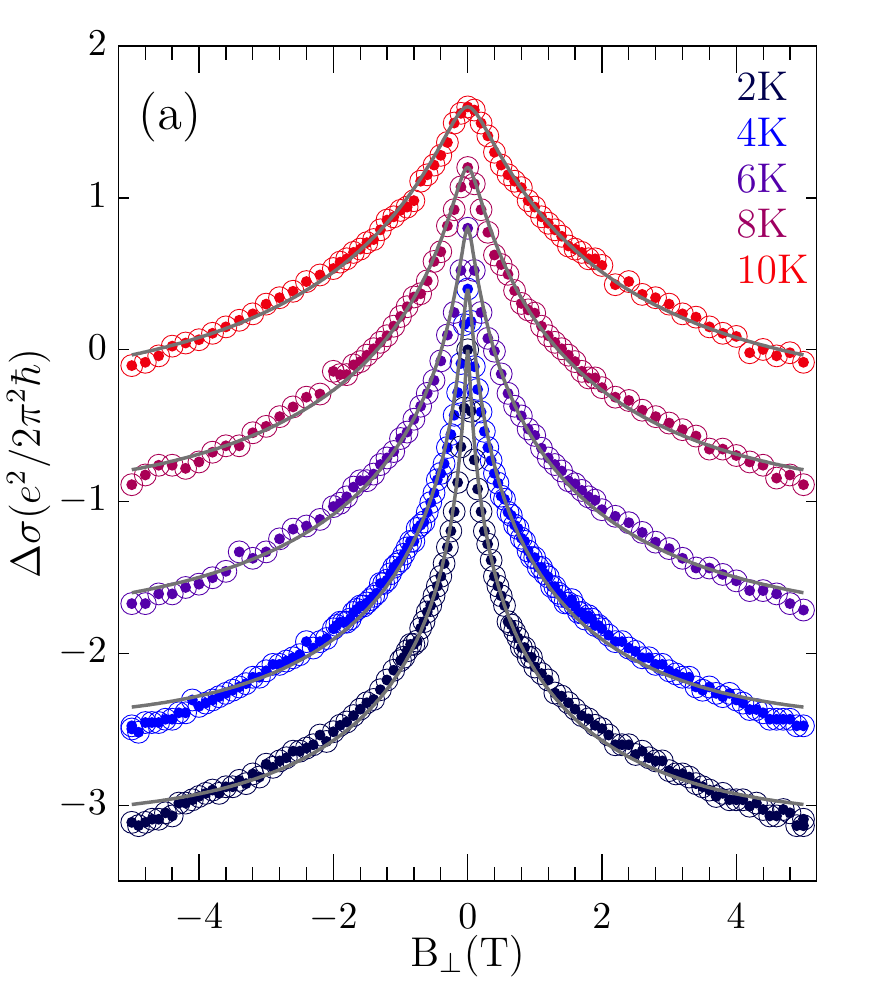}\\
	\includegraphics[width=7.6cm,height=6cm]{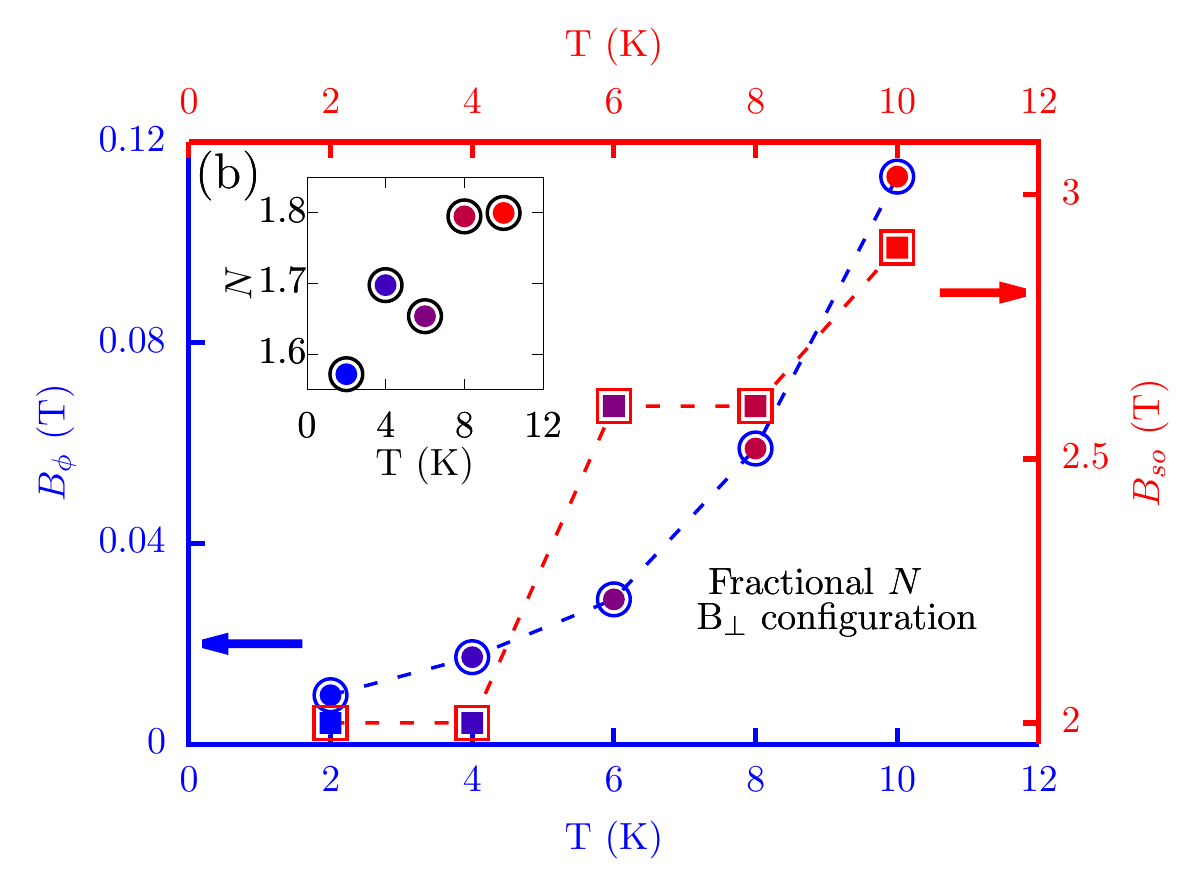}
	\end{center}
	\caption{\small \textcolor{black}{{(a) Magnetoconductance $\Delta \sigma (\B_{\perp})$ is fitted with HLN equation at different temperature for 16 nm Pt film (solid gray color lines represent fitting of  Eq.(\ref{Eq_HLN_frac_channel})) (vertically	offset to plots are given for clarity).  (b) Extracted $B_{\phi}~B_{so}$ are illustrated  at different temperatures, inset displays the fractional value of $N$ at various temperatures.}}
		\label{fig_16nm_sigma_HLN_frac}}
\end{figure}


The magnetotransport measurement is a powerful tool to extract different types of microscopic scattering lengths  ($i.e.$  spin-orbit, inelastic scattering) by exploiting the quantum interference  originated correction (WAL/WL) to sheet conductance.  Quantum interference effect (WL/WAL) manifests more prominently with  the reduction of dimension.
%
%
%
\begin{figure}[t!]
	\begin{center}
		\includegraphics[width=7cm,height=7 cm]{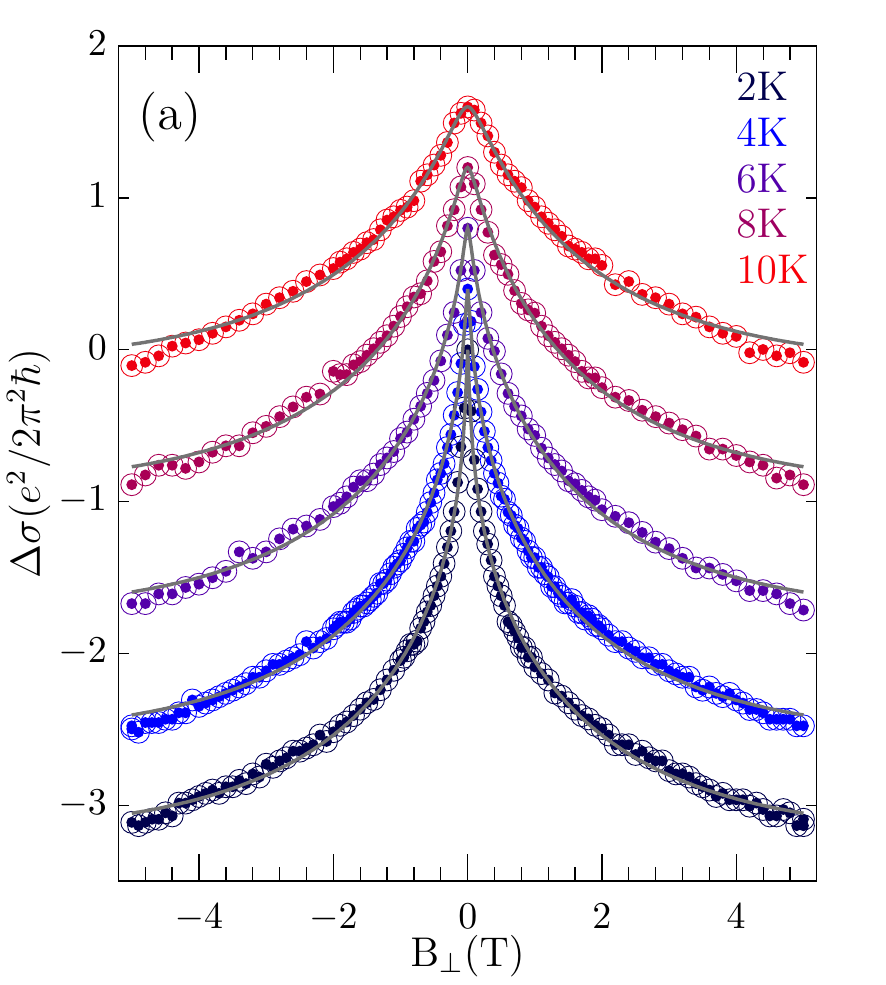}\\
		\includegraphics[width=7.6cm,height=6cm]{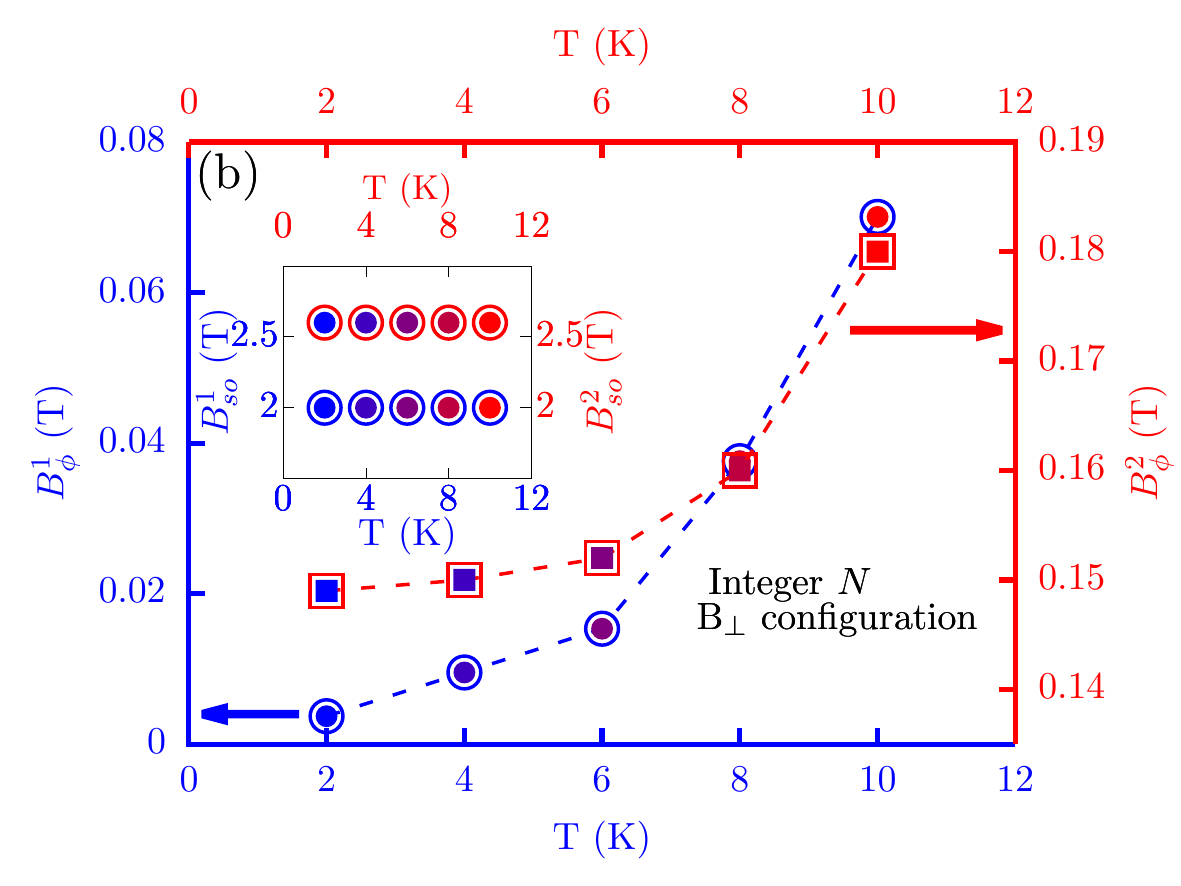}
	\end{center}
	\caption{\small \textcolor{black}{{(a) Magnetoconductance $\Delta \sigma (\B_{\perp})$ is fitted with HLN equation at different temperature (solid gray color lines represent fitting of  Eq.(\ref{Eq_HLN_multi})) for 16 nm Pt film (vertically	offset to plots are given for clarity). (b) Extracted $B^{n}_{\phi}$ for two different independent channels ($n=1,2$)  are illustrated  at different temperatures,  inset displays the  value of $B^{n}_{so}$ at various temperatures.}}
		\label{fig_16nm_sigma_HLN_multi}}
\end{figure}
%
For quasi-2D system (where  phase coherent length ($l_{\phi}$) is greater than film thickness ($t$)) with $N$ independent conducting channels, the  variation of sheet conductance ($\Delta\sigma(\B_{\perp})= \sigma(\B_{\perp}) -\sigma(0) $) with  external perpendicular magnetic field ($B_{\perp}$)  is described by Hikami-Larkin-Nagaoka (HLN) equation which can be expressed as \cite{hikami1980spin,al1981magnetoresistance,dirac_1,dirac_2,WAL_WL_2},

\begin{equation}%
\begin{split}
\Delta\sigma(\B _{\perp}) &=  - \dfrac{e^{2}}{2\pi^{2}\hslash} \sum_{n=1}^{N}  \bigg[  \bigg\{\psi(1/2 + B^{n}_{e}/\B_{\perp}) + \ln(\B_{\perp}/B^{n}_{e}) \bigg\} \\
& +\frac{1}{2}\bigg\{\psi(1/2 + B^{n}_{\phi}/\B_{\perp}) + \ln(\B_{\perp}/B^{n}_{\phi})\bigg\} \\ 
& -\frac{3}{2} \bigg\{\psi(1/2 + \frac{B^{n}_{\phi} + B^{n}_{so}}{\B_{\perp}}) + \ln(\frac{\B_{\perp}}{B^{n}_{\phi} + B^{n}_{so}})\bigg\} \bigg] 
\end{split}
\label{Eq_HLN_multi}
\end{equation}
where $\psi(x)$ is the digamma function,  $B^{n}_{e}, B^{n}_{\phi} $ and $ B^{n}_{so}$   correspond to characteristic magnetic fields of $n$th channel  which is  related with scattering lengths  ($l_{e}, l_{\phi}$ and $ l_{so}$) as
$B_{i} = \hslash /4el^{2}_{i}$, where $l_{e},~l_{\phi}$ and $~l_{so}~$ denote elastic, phase coherence and spin-orbit scattering lengths respectively  and $B_{e}$ is determined from semi-classical approximation and detailed discussion is given in supplemental material \cite{Suppl_Material}. 

\begin{figure}[t!]
	\begin{center}
		\includegraphics[width=7cm,height=7 cm]{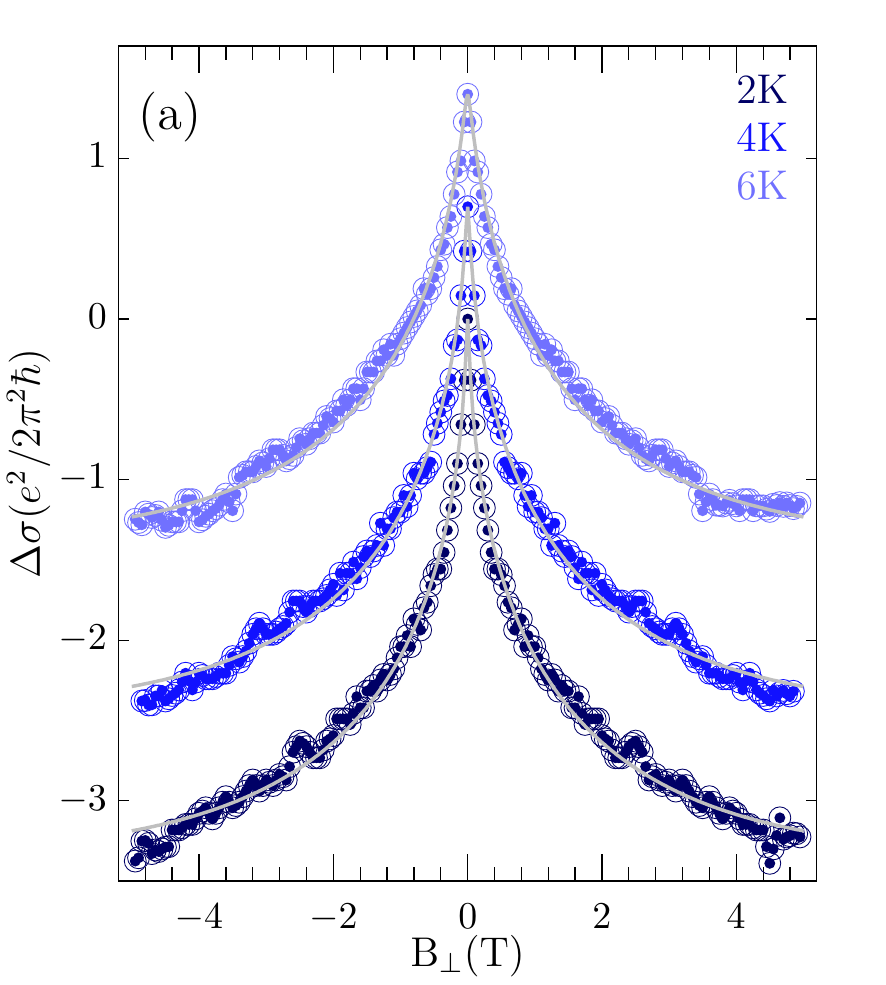}\\
		\includegraphics[width=7.6cm,height=6cm]{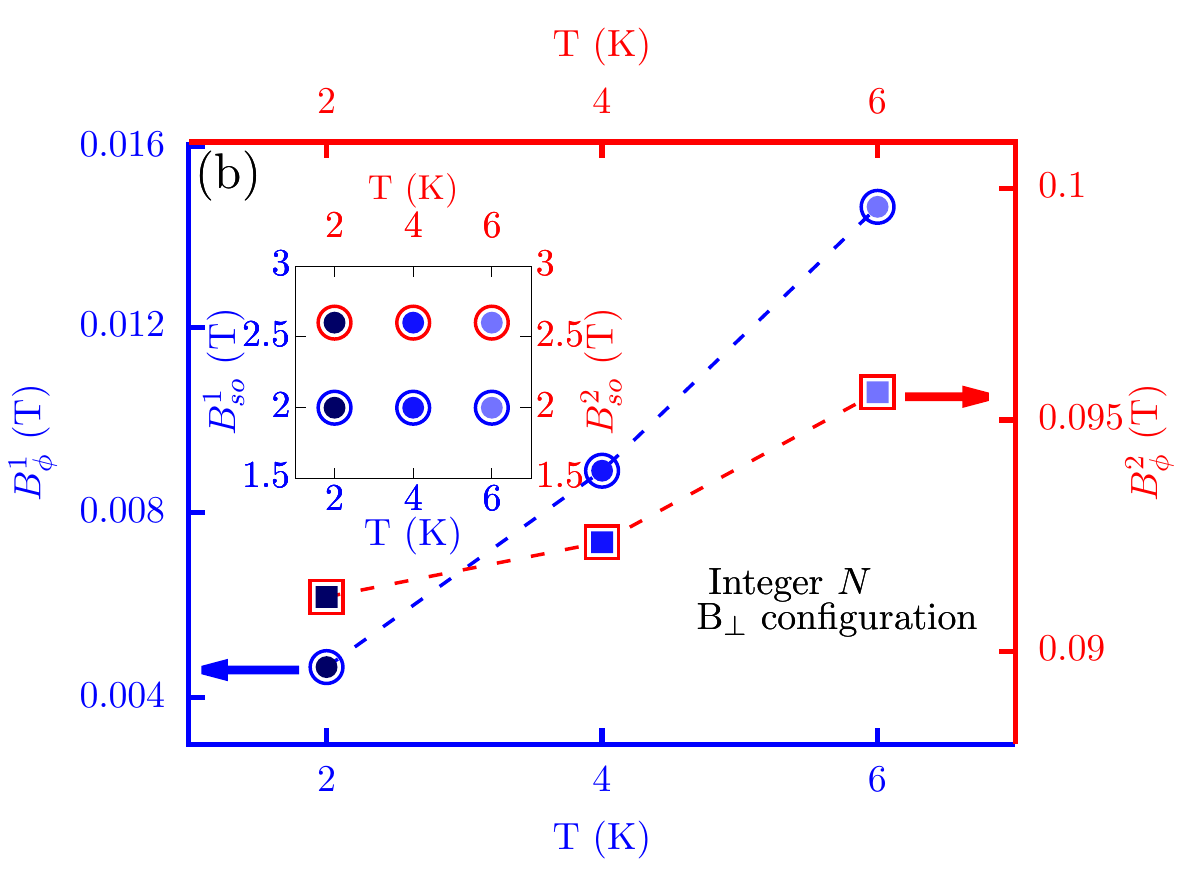}
	\end{center}
	\caption{\small \textcolor{black}{{(a) Magnetoconductance $\Delta \sigma (\B_{\perp})$ is fitted with HLN equation at different temperature (solid gray color lines represent fitting of  Eq.(\ref{Eq_HLN_multi})) for 23 nm Pt film (vertically	offset to plots are given for clarity). \\ (b) Extracted $B^{n}_{\phi}$ for two different independent channels ($n=1,2$)  are illustrated  at different temperature,  inset displays the  value of $B^{n}_{so}$ at various temperatures.}}
		\label{fig_23nm_sigma_HLN_multi}}
\end{figure}
%
If independent channels are very much similar to each other, then the characteristic magnetic fields $B^{n}_{i}$ corresponding to each channel are nearly equal and as a result  HLN Eq.(\ref{Eq_HLN_multi}) for $N$ channels  becomes 
\begin{equation}%
\begin{split}
\Delta\sigma(\B _{\perp}) &=  -N \dfrac{e^{2}}{2\pi^{2}\hslash} \bigg[  \bigg\{\psi(1/2 + B_{e}/\B_{\perp}) + \ln(\B_{\perp}/B_{e}) \bigg\} \\
& +\frac{1}{2}\bigg\{\psi(1/2 + B_{\phi}/\B_{\perp}) + \ln(\B_{\perp}/B_{\phi})\bigg\} \\ 
& -\frac{3}{2} \bigg\{\psi(1/2 + \frac{B_{\phi} + B_{so}}{\B_{\perp}}) + \ln(\frac{\B_{\perp}}{B_{\phi} + B_{so}})\bigg\} \bigg].
\end{split}
\label{Eq_HLN_frac_channel}
\end{equation}
\begin{figure}[t!]
	\begin{center}
		\includegraphics[width=7cm,height=7 cm]{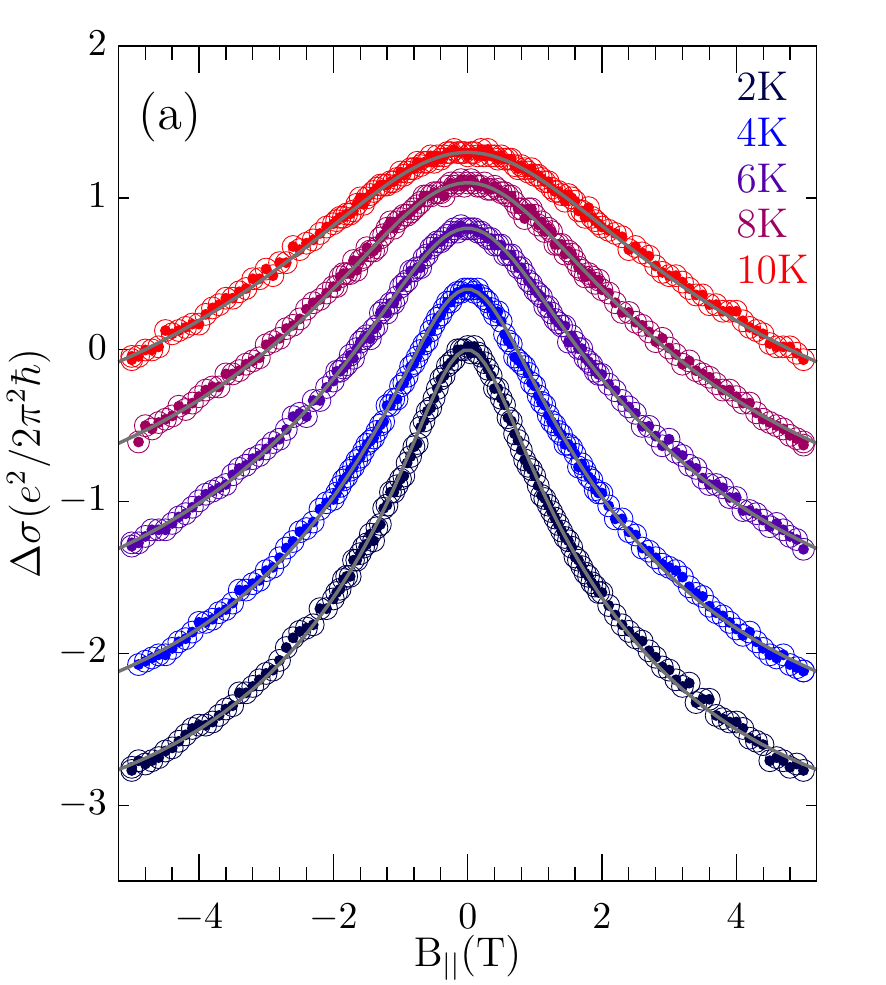}\\
		\includegraphics[width=7.6cm,height=6cm]{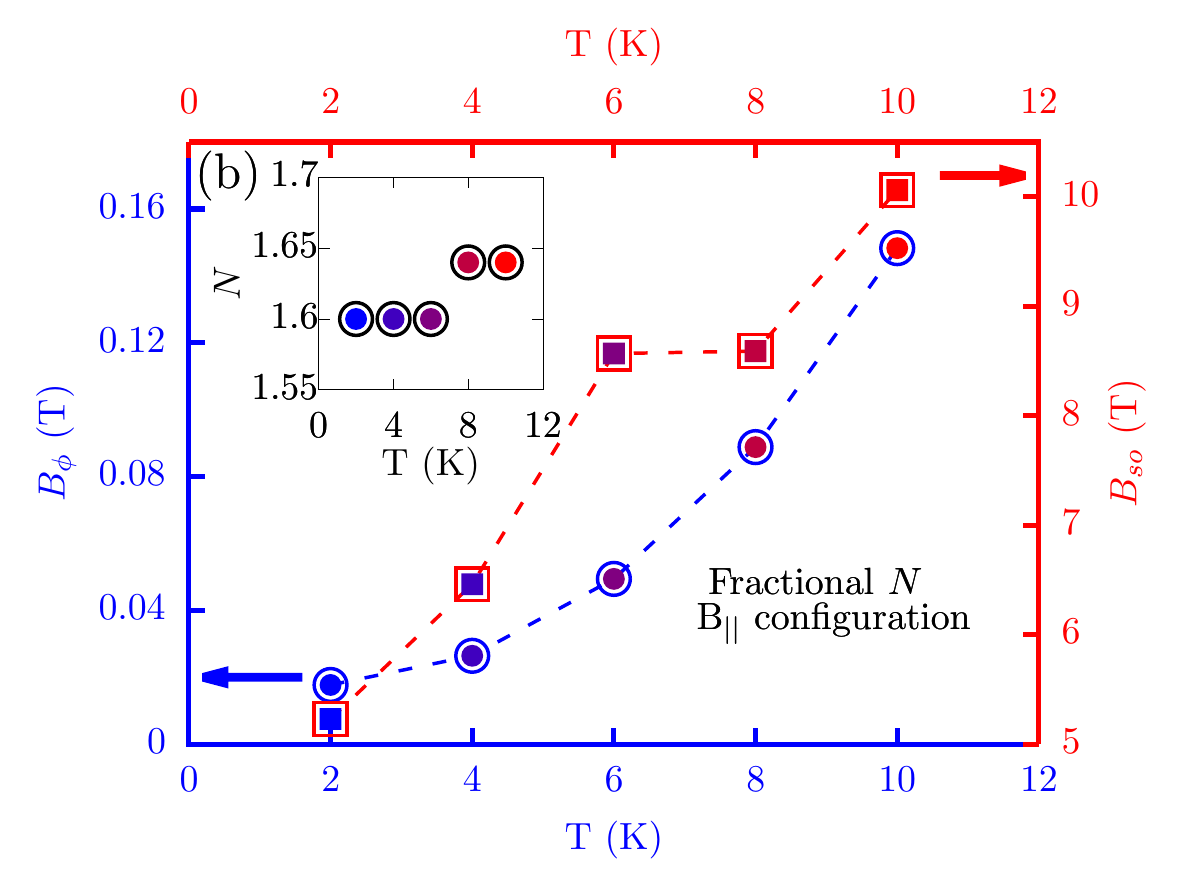}
	\end{center}
	\caption{\small \textcolor{black}{{(a) Magnetoconductance $\Delta \sigma (\B_{||})$ is fitted with HLN equation at different temperatures for 16 nm Pt film (solid gray color lines represent fitting of  Eq.(\ref{Eq_HLN_frac_channel})) (vertically	offset to plots are given for clarity).  (b) Extracted $B_{\phi}~B_{so}$ are illustrated  at different temperature,  inset displays the fractional value of $N$ at various temperatures.}}
		\label{fig_16nm_sigma_HLN_frac_parallel}}
\end{figure}

Fig.\ref{fig_16nm_sigma_HLN_frac}(a) shows the sheet conductance variation with  perpendicular magnetic field for 16 nm thick film at different temperature.
The presence of sharp cusp at lower temperature indicates the WAL effect. However, the sharpness  disappears rapidly with the increment of temperature.  
Experimental data $\Delta \sigma(B_{\perp})$ are fitted well with Eq.(\ref{Eq_HLN_frac_channel}), nevertheless the fitting  provides fractional value of $N$ which can not be anticipated by considering independent conducting channels. The fitting and their corresponding best fitting parameters ($B_{\phi},B_{so},N$ ) are illustrated in Fig.\ref{fig_16nm_sigma_HLN_frac}(a) and \ref{fig_16nm_sigma_HLN_frac}(b) respectively.
The fractional value of $N$ can appear due to following reasons, (i) presence of weak but non-negligible inter-orbital scattering  among channels \cite{frac_N,nakamura2020robust} and 
(ii)  channels posses different set of $B^{n}_{e},~B^{n}_{so},~B^{n}_{\phi}~~$.

\begin{figure}[t!]
	\begin{center}
		\includegraphics[width=7cm,height=7 cm]{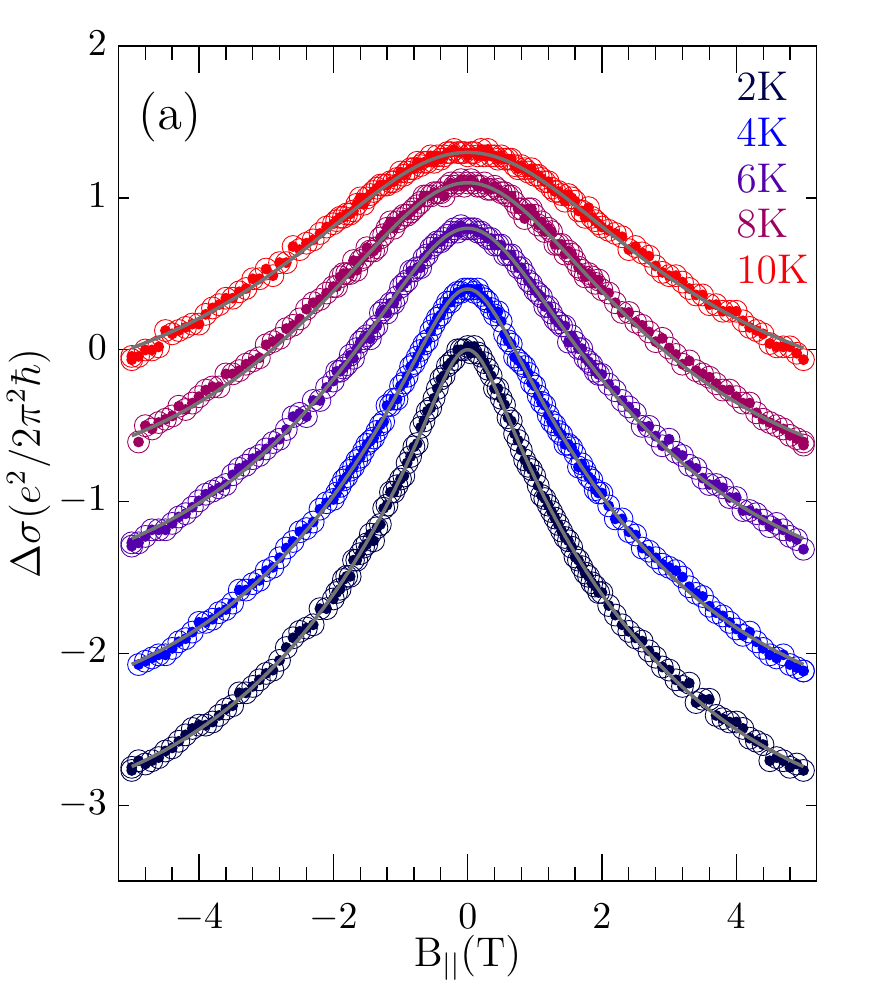}\\
		\includegraphics[width=7.6cm,height=6cm]{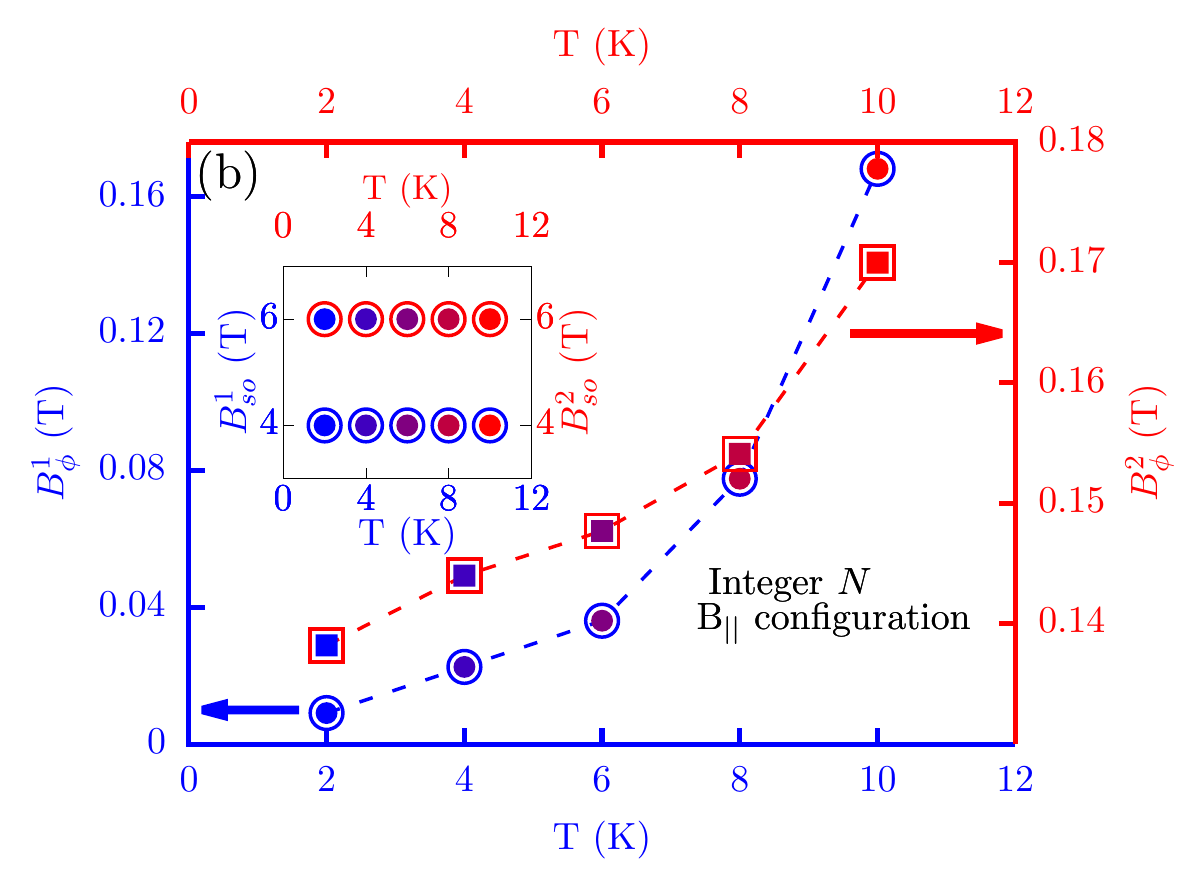}
	\end{center}
	\caption{\small \textcolor{black}{{(a) Magnetoconductance $\Delta \sigma (\B_{||})$ is fitted with HLN equation at different temperatures (solid gray color lines represent fitting of  Eq.(\ref{Eq_HLN_multi})) for 16 nm Pt film (vertically	offset to plots are given for clarity). (b) Extracted $B^{n}_{\phi}$ for two different independent channels ($n=1,2$)  are illustrated  at different temperatures,  inset displays the  value of $B^{n}_{so}$ at various temperatures.}}
	\label{fig_16nm_sigma_HLN_multi_parallel}}
\end{figure}

To obtain more meaningful insight about the conducting channels in Pt thin films, the magnetoconductance data is further analyzed  by considering more general equation (Eq.\ref{Eq_HLN_multi}) with the independent channels  having different set of $B^{n}_{e},~B^{n}_{so},~B^{n}_{\phi}$. The fitting is shown in Fig.\ref{fig_16nm_sigma_HLN_multi}(a) and the best fitting parameters ($B^{1}_{\phi},B^{2}_{\phi},B^{1}_{so},B^{2}_{so}$) are shown in Fig.\ref{fig_16nm_sigma_HLN_multi}(b) (for 16 nm thick Pt film). From the fitting, we obtain $N=2$ and the phase coherence magnetic field ($B^{1}_{\phi}\sim 0.004$ T at 2 K) for one channel to be  higher than other ($B^{2}_{\phi} \sim 0.15$ T at 2 K). Further, spin-orbit scattering magnetic field $B^{1}_{so} \sim 2$ T ,$~B^{2}_{so} \sim 2.5$ T are found to be very much higher than both $B^{1}_{\phi},~B^{2}_{\phi}$.
%
The obtained fitting parameters provide the following physical interpretations:
(i) the   Pt  metal possesses large intrinsic spin-orbit interaction $B_{so}\sim 2$ T (in perpendicular configuration with integer number channels $N=2$),  
(ii) presence of two independent conducting channels  in Pt thin film, and  (iii)  the variation of $B^{1}_{\phi}$ with  temperature   is very much prominent in comparison with the $B^{2}_{\phi}$ and $B^{1}_{\phi} < B^{2}_{\phi}$. The prominent variation of  $B^{1}_{\phi}~vs~T$ for one channel indicates that  the inelastic scattering process  is dominated by $e$-$e$ scattering (Sec:\ref{rho_T} ). However, the other channel with weakly temperature dependent  inelastic scattering process (at low temperature regime $2<T<6$ K) indicates that significant  amount of impurity scattering is present. 
To corroborate further, we analyzed the same for 23 nm thick Pt film and we found that it also follows similar behavior. The fitting and obtained parameters are illustrated in Fig.\ref{fig_23nm_sigma_HLN_multi}(a), \ref{fig_23nm_sigma_HLN_multi}(b).    
%
\begin{figure}[t!]
	\begin{center}
		\includegraphics[width=7cm,height=7 cm]{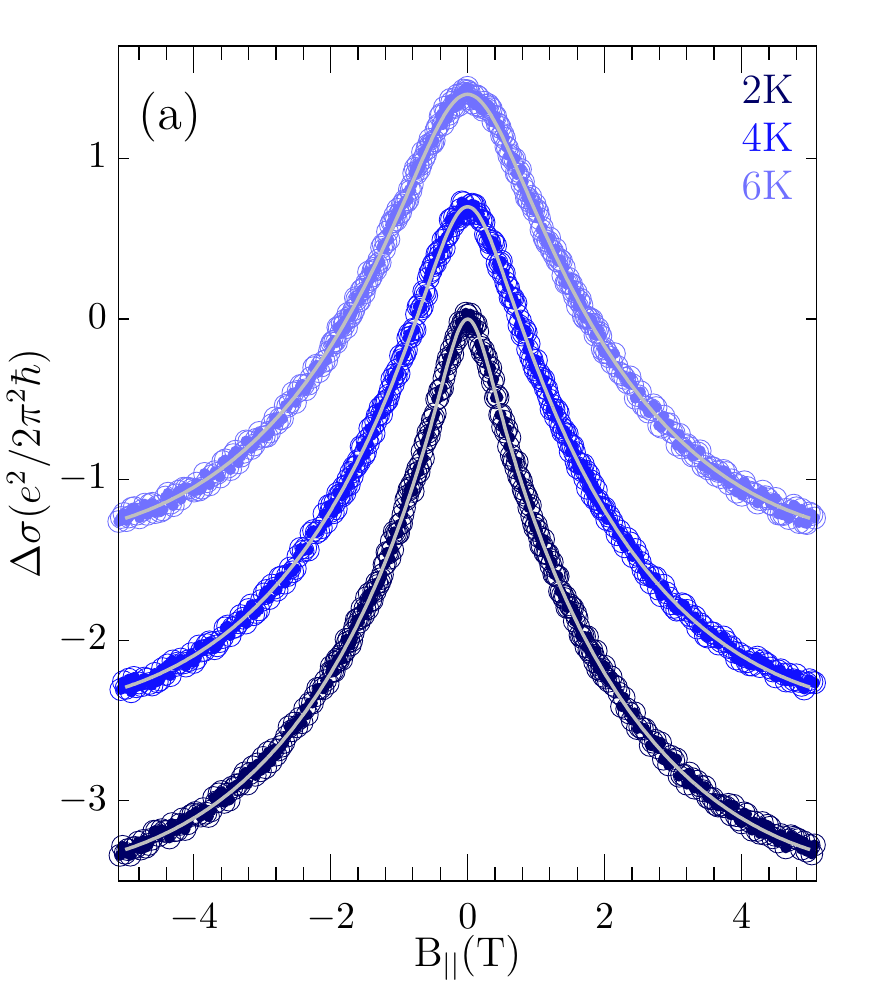}\\
		\includegraphics[width=7.6cm,height=6cm]{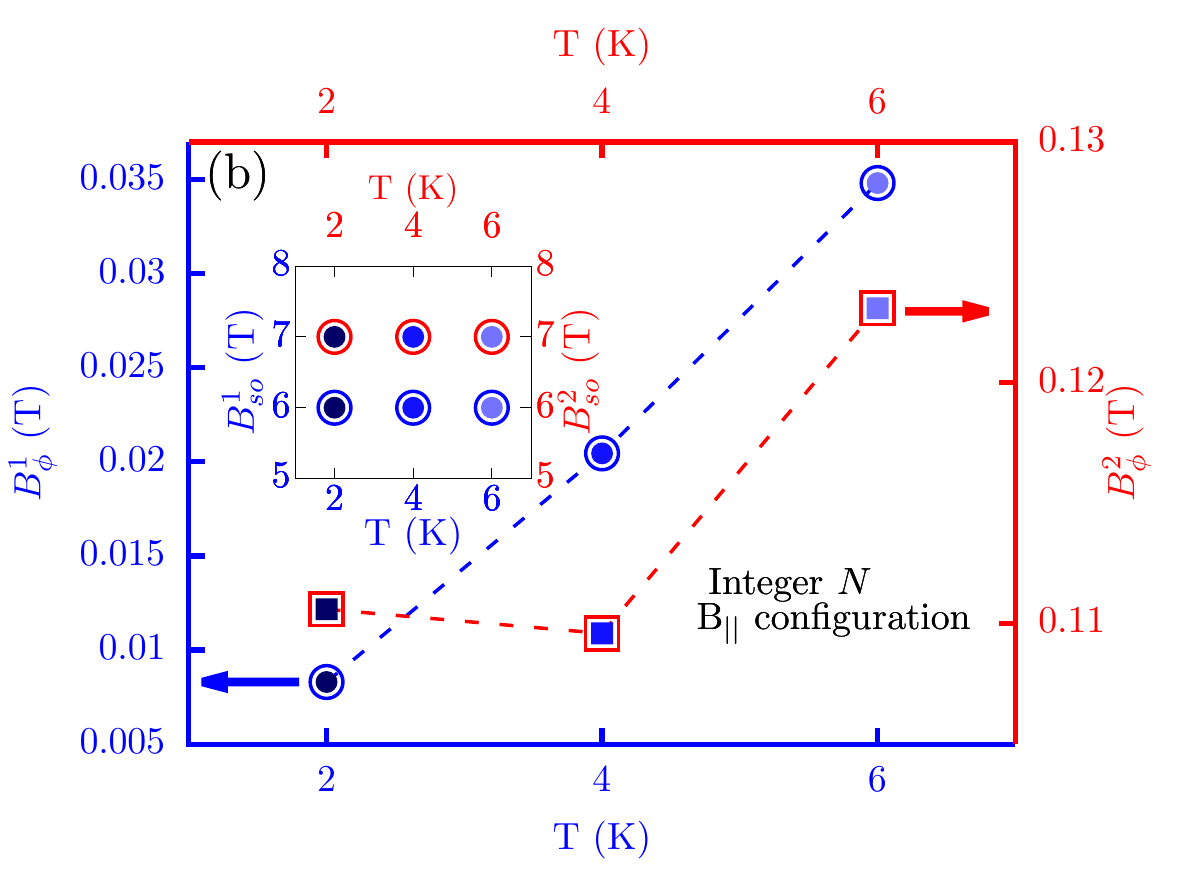}
	\end{center}
	\caption{\small \textcolor{black}{{(a) Magnetoconductance $\Delta \sigma (\B_{||})$ is fitted with HLN equation at different temperatures (solid gray color lines represent fitting of  Eq.(\ref{Eq_HLN_multi})) for 23 nm Pt film (vertically	offset to plots are given for clarity).  (b) Extracted $B^{n}_{\phi}$ for two different independent channels ($n=1,2$)  are illustrated  at different temperatures, inset displays the  value of $B^{n}_{so}$ at various temperatures.}}
		\label{fig_23nm_sigma_HLN_multi_parallel}}
\end{figure}


 

\subsection{Parallel Magnetic field ($B_{||}$)}
\label{B_parallel}
In quasi-2D system, itinerant electrons can diffuse along film thickness by satisfying the restriction of quantum mechanical boundary conditions \cite{al1981magnetoresistance}.   Hence, quantum interference effect gets  influenced even  in parallel magnetic field configuration. Quantum interference originated magnetoconductance correction in $B_{||}$ configuration  for $N$ independent channels is given by\cite{al1981magnetoresistance,PhysRevB.32.2190,PhysRevB.29.735,PhysRevB.32.6319,PhysRevB.86.205302},

\begin{equation}%
\begin{split}
\Delta\sigma(\B _{||}) &=  \dfrac{e^{2}}{2\pi^{2}\hslash} \sum_{n=1}^{N}  \bigg[   \frac{3}{2} \ln \bigg( 1 + \frac{\B^{2}_{||}}{B_{t}.(B^{n}_{so} + B^{n}_{\phi} )  }  \bigg)\\
& -   \frac{1}{2} \ln \bigg( 1 + \frac{\B^{2}_{||}}{B_{t} .B^{n}_{\phi}   }  \bigg)         \bigg] 
\end{split}
\label{Eq_HLN_multi_parallel}
\end{equation}
where, $B_{t}$ is defined as, $B_{t} = 12\hbar /et^{2}$ and $t$ is film thickness.

For simplicity, if we assume all independent channels to have  equal characteristic magnetic filed then Eq.(\ref{Eq_HLN_multi_parallel}) gets modified as
\begin{equation}%
\begin{split}
\Delta\sigma(\B _{||}) &= N. \dfrac{e^{2}}{2\pi^{2}\hslash}   \bigg[   \frac{3}{2} \ln \bigg( 1 + \frac{\B^{2}_{||}}{B_{t}.(B_{so} + B_{\phi} )  }  \bigg)\\
& -   \frac{1}{2} \ln \bigg( 1 + \frac{\B^{2}_{||}}{ B_{t} .B_{\phi}   }  \bigg)         \bigg]. 
\end{split}
\label{Eq_HLN_multi_frac}
\end{equation}

Experimental data $\Delta \sigma(B_{||})$ are fitted well with Eq.(\ref{Eq_HLN_multi_frac}) (Fig.\ref{fig_16nm_sigma_HLN_frac_parallel}(a)), but the fitting  provides fractional value of $N$. The extracted parameters are illustrated in Fig.\ref{fig_16nm_sigma_HLN_frac_parallel}(b) for 16 nm thick film. As, previously we have discussed that  number of independent conducting channel can not be a fractional value,  different channels must have different characteristic magnetic fields.  Therefore, we  analyzed  $ \Delta\sigma(\B _{||})$ considering  the independent channels to have different set of $~B^{n}_{so},~B^{n}_{\phi}$  (Eq.(\ref{Eq_HLN_multi_parallel})) which is shown in Fig.\ref{fig_16nm_sigma_HLN_multi_parallel}.(a) (for 16 nm thick Pt film), Fig.\ref{fig_23nm_sigma_HLN_multi_parallel}.(a) (for 23 nm thick Pt film) and the best fitting parameters are shown in Fig.\ref{fig_16nm_sigma_HLN_multi_parallel}.(b) (for 16 nm thick Pt film), Fig.\ref{fig_23nm_sigma_HLN_multi_parallel} (b) (for 23 nm thick Pt film) 

In parallel magnetic field configuration, the extracted values of $B^{1}_{\phi}$ (by fitting with Eq.\ref{Eq_HLN_multi_parallel}) are found to be petty close to $B^{1}_{\phi}$ obtained (by fitting with Eq.\ref{Eq_HLN_multi}) in  perpendicular  magnetic field configuration  and follows  very much similar behaviour with temperature. However, the obtained $B_{so}$ are very much different for perpendicular and parallel configuration. It confirms  the  anisotropic nature of  the spin-orbit scattering potential in Pt thin films.

\begin{figure}[t!]
	\begin{center}
		\includegraphics[width=7cm,height=5.6 cm]{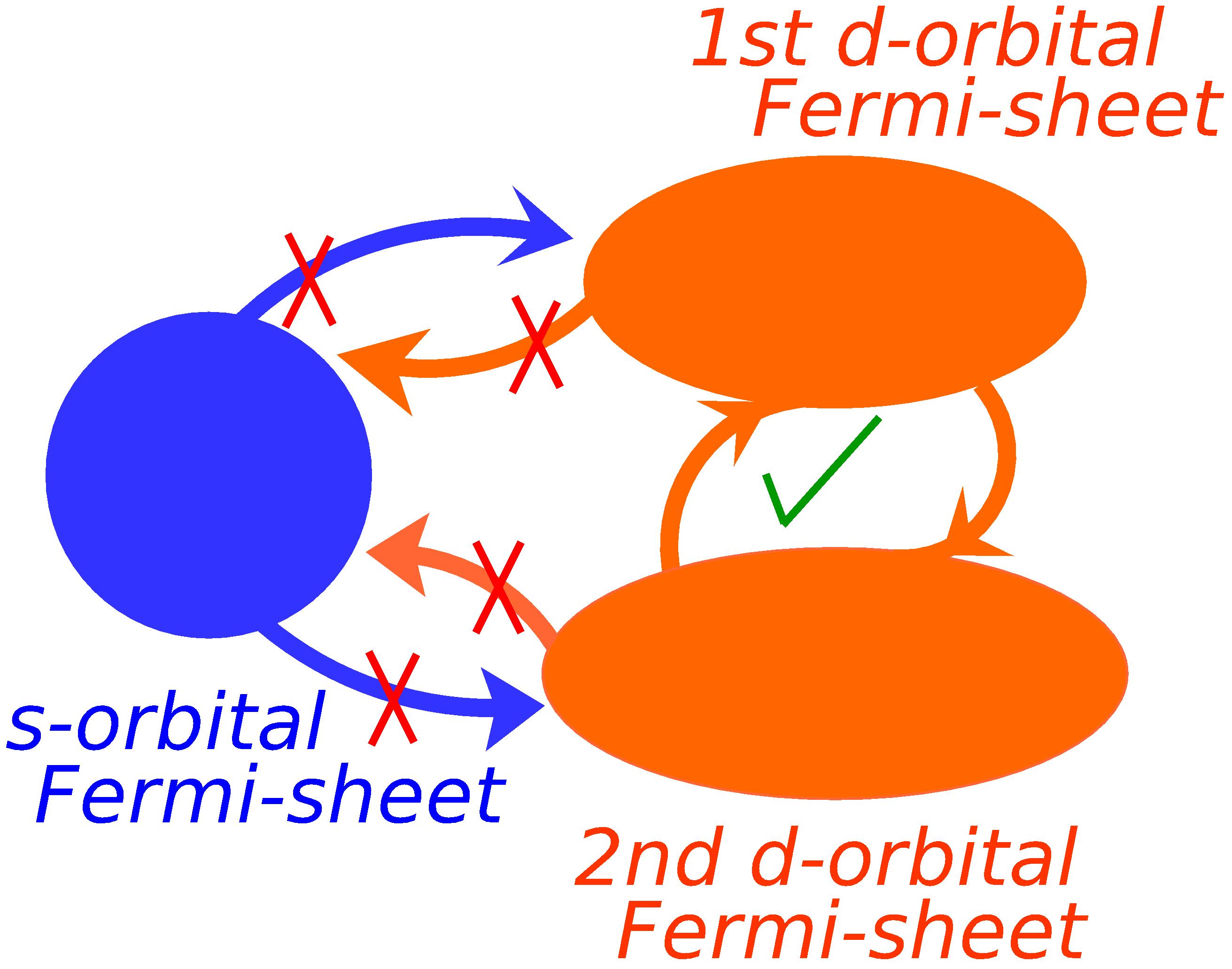}
	\end{center}
	\caption{\small \textcolor{black}{{The schematic diagram for theoretically predicated three Fermi-sheets one derived from $s$-orbital (blue color) and other two derived $d$-orbital (orange color).  The possible interaction  among them is denoted by arrows (allowed and forbidden interaction  is marked by $\times,~\checkmark$ respectively). Interconnected two $d$-orbitals leads to one effective conducting channel and other one is from $s$-orbitals}}
		\label{fig_schematic}}
\end{figure}

\section{$\text{Summary and Conclusions}$}


It was predicted from a theoretical band structure calculation that  Pt metal (electronic configuration 5$d^{9}$6$s^{1}$) has three sheets of Fermi surfaces which are comprised of  $s$-band and  two from $d$-band \cite{Pt_dft}.  We realised  the presence of two independent conduction channels from our magneto-transport analysis in both perpendicular and parallel magnetic configurations. This apparent discrepancy (theoretically three channels, experimentally two channels) leads to an insightful physics of Pt thin film in presence of  weak disorder scattering which is unavoidable in real system. The two $d$-band channels are mixed up with each other due to their similar orbital nature and in presence of intermixing scattering, they will act as an equivalent single channel.  On the contrary, intermixing scattering among   $s,~d$-band channels is very weak due to different orbital symmetry. As a result effectively there will be  two independent conducting channels which is illustrated in Fig.\ref{fig_schematic}.  
Secondly, we have found that the  characteristic magnetic field for coherence phase breaking scattering ($B^{2}_{\phi}$) corresponding to one channel  shows a weak temperature dependence and this intriguing phenomena could  arise  very often in a system with significant impurity scattering. However, $B^{1}_{\phi}$  for other  channel  shows a prominent variation with temperature, indicating the  the presence of  effectively weak impurity scattering. The crucial point is that the same amount of disorder in a system can act  differently for different conduction channels depending upon  it's orbital symmetry.  The bands derived from more anisotropic orbitals  ($e.g.~d,~p$ orbital) are very much sensitive to minute amount of disorder \cite{mott1968conduction,mott1969conduction} which  can give rise to very high inelastic scattering in comparison to conducting channel made out of symmetric  orbitals ($e.g. ~s$ orbital). To conclude, the symmetry of  orbitals involved in conducting channels, presence of EEI and anisotropic spin-orbit interaction are revealed from our magnetoconductance study in Pt thin films.

\vspace{6 mm}
\begin{acknowledgements}
SJ  and DS  acknowledge the financial support from Max Planck partner group. DS  acknowledges the funding support from SERB, Govt. of India.  SJ  would like to thank R. Kumar for discussions.
\end{acknowledgements}
\bibliography{Pt.bib}

\begin{thebibliography}{69}
\bibitem [{Td()}]{Td}%
  \BibitemOpen
  \href@noop {} {\bibinfo  {journal} {Function ln(x) is not defined for
  dimensioned quantities. Therefore, to make the argument dimensionless, we use
  ln(T/T$_{d}$) as the x-axis label of Fig.2 with T$_{d}=$1 K}\ }\BibitemShut
  {NoStop}%
\bibitem [{fra()}]{frac_N}%
  \BibitemOpen
\bibfield  {journal} {  }\href@noop {} {\bibinfo  {journal} {Two extereme
  situations $i.e.$ in presence (absence) of strong inter-orbital scattering
  among N channels gives rise to effective single channel (N channels) for
  magnetotransport phenomena. Therefore, intutively non-integer value of
  independent channel may arise for intermediate (small but non-negligible)
  inter-orbital scattering strength}\ }\BibitemShut {NoStop}%
\end{thebibliography}
\end{document}